\documentclass[conference]{IEEEtran}
\usepackage[T1]{fontenc}

\makeatletter
\long\def\@makecaption#1#2{\ifx\@captype\@IEEEtablestring%
\footnotesize\begin{center}{\normalfont\footnotesize #1}\\
{\normalfont\footnotesize\scshape #2}\end{center}%
\@IEEEtablecaptionsepspace
\else
\@IEEEfigurecaptionsepspace
\setbox\@tempboxa\hbox{\normalfont\footnotesize {#1.}~~ #2}%
\ifdim \wd\@tempboxa >\hsize%
\setbox\@tempboxa\hbox{\normalfont\footnotesize {#1.}~~ }%
\parbox[t]{\hsize}{\normalfont\footnotesize \noindent\unhbox\@tempboxa#2}%
\else
\hbox to\hsize{\normalfont\footnotesize\hfil\box\@tempboxa\hfil}\fi\fi}
\makeatother

\usepackage[english]{babel}
\usepackage{calc}
\usepackage{amsmath,amssymb,amsfonts}
\usepackage{bbm}
\usepackage[noadjust]{cite}
\usepackage[tight,footnotesize]{subfigure}
\usepackage{graphicx}
\usepackage{algorithm}
\usepackage{algpseudocode}
\usepackage{epstopdf}
\usepackage{multirow}
\usepackage{stfloats}

\usepackage{fancyhdr}
\newlength{\EndIterLen}
\algblockdefx[Init]{Init}{EndInit}{\textbf{Initialization}}{}
\algblockdefx[Iter]{Iter}{EndIter}{\textbf{Iteration Loop}}%
{\textbf{End Iteration Loop}\settowidth{\EndIterLen}{\textbf{End Iteration Loop}}}
\algblockdefx[MyFor]{MyFor}{EndMyFor}[1]{\textbf{for all} #1 \textbf{do}}{}
\algblockdefx[MyIf]{MyIf}{EndMyIf}[1]{\textbf{if} #1 \textbf{then}}{}
\algtext*{EndInit}
\algtext*{EndMyFor}
\algtext*{EndMyIf}

\newcommand{\vect}[1]{\underline{\boldsymbol #1}}
\newcommand{\defeq}{\stackrel{\Delta}{=}}

\usepackage{xcolor}
\newcommand{\m}{\textcolor{gray}{-1}}

\IEEEoverridecommandlockouts

\begin{document}


%
\title{Gradient Descent Bit-Flipping Decoding with Momentum\vspace*{-2mm}}
%
%
%



\author{Valentin Savin, CEA-LETI, Universit\'e Grenoble Alpes,  France (valentin.savin@cea.fr)\vspace*{-1mm}%
\thanks{This work was partially supported by the French Agence Nationale de la Recherche (ANR), under grant number ANR-15-CE25-0006 (NAND project), and by the IPCEI programme of the European Commission, and the French Minister of Finance through DGE organization.}}

%



\maketitle

\thispagestyle{fancy}

\begin{abstract}
In this paper, we propose a Gradient Descent Bit-Flipping (GDBF) decoding with momentum, which considers past updates to provide inertia to the decoding process. We show that GDBF or randomized GDBF decoders with momentum may closely approach  the floating-point Belief-Propagation decoding performance, and even outperform it in the error-floor region, especially for graphs with high  connectivity degree. 
\end{abstract}


%

%

\section{Introduction}
Bit-Flipping (BF) is the simplest form of iterative decoding for codes defined by bipartite graphs, requiring only one $1$-bit message per graph node, and per iteration. Compared to  message-passing (MP) decoders -- exchanging multi-bit {\em extrinsic messages} along the graph edges -- BF yields a significant reduction of the computational time and space complexity. As one may expect, this complexity reduction comes at the price of a significant degradation of the error correction performance. BF decoding was already introduced by Gallager in his seminal paper \cite{gallager1963low}, as a very first example of simple, but poor performance approach to iterative decoding. 
Probabilistic decoding came then into play, latter reformulated in terms of Belief-Propagation (BP),   paving the way for many  MP decoders subsequently proposed in the literature. 
In a sense, the original BF decoding has been relegated to a secondary (and for many years almost inexistent) role.

However, the emergence of massive data rate communications over the last few years, has motivated a renew of interest for BF or BF-based decoding, seen as a possible approach  towards meeting the very stringent requirements in terms of throughput/latency and energy efficiency. It is worth mentioning here the reformulation of BF decoding  as a gradient descent solution to an optimization problem, proposed in \cite{wadayama2008gradient}. The corresponding decoding algorithm, referred to as Gradient Descent Bit-Flipping (GDBF), is appealing in practice due to its simplicity, and since it answers an optimization problem that may, in principle, yield a solution to the ML decoding problem (see Section~\ref{sec:randomized_gdbf}). However, its main drawback is that it suffers from many local optimum traps, due to the high non-linearity of the objective function. Therefore, even if GDBF performs significantly better than the original BF decoding proposed by Gallager, or other variants of the BF decoding proposed in the literature, its performance is still behind that of more powerful MP decoders, such as BP or Min-Sum (MS).

The current solution to further improving the GDBF decoding performance relies on randomization. To some extent, the approach is reminiscent of noisy decoders,  intensively investigated in the literature over the last years \cite{varshney2011performance, yazdi2013gallager, kameni2015density, dupraz2015analysis}. 
The difference between noisy and randomized decoders is that in the former case, noise is an external aggression that perturbs the decoding operations, while in the latter case, noise (or strictly speaking, randomness) has desired statistical properties and is an integral part of decoding operations.  Moreover, it has been recently shown in the literature that randomness allows GDBF decoders to escape from local optima, thus improving their error correction performance. Randomized versions of the GDBF decoder -- {\em e.g.}, Probabilistic GDBF (PGDBF)  \cite{rasheed2014fault} and  Noisy GDBF (NGDBF)  \cite{sundararajan2014noisy} -- have been shown to considerably narrow the gap to more powerful MP decoders. It is worth noting that both PGDBF and NGDBF introduce randomness in the bit flipping rule, and they carry strong similarities to each other. The PGDBF is however a hard-decision decoding algorithm, thus it mainly applies to the binary symmetric channel (BSC) model, while NGDBF applies to more general soft-output channels.  

In this paper, we investigate alternative approaches to further improve the GDBF decoding performance, inspired by techniques used in gradient descent optimization. 
%
%
%
We propose a GDBF decoding with momentum, which considers past updates to provide inertia to the decoding process. We show that GDBF or randomized GDBF decoders with momentum may closely approach the floating-point BP decoding performance, and even outperform it in the error-floor region, especially for graphs with high degree of connectivity. 

\section{Randomized GDBF Decoders}\label{sec:randomized_gdbf}
We consider an LDPC code defined by a bipartite (Tanner) graph $H$, with $N$ variable-nodes (corresponding to coded bits) and $M$ check-nodes (corresponding to parity-check equations). We denote by $H(n)$ the set of check-nodes connected to a variable-node $n=1,\dots,N$, and by $H(m)$ the set of variable-nodes connected to a check-node $m=1,\dots,M$. We assume either a binary symmetric channel (BSC), or a binary-input additive white Gaussian noise (AWGN) channel. For simplicity with shall assume that in both cases the input alphabet of the channel is $\{-1, +1\}$, thus transmitted codewords take values in $\{-1, +1\}^N$ rather than $\{0, 1\}^N$.
The maximum likelihood (ML) decoding is equivalent to finding a codeword $\vect{x} = (x_1,\dots,x_N)$, having the maximum correlation with the received word $\vect{y} = (y_1,\dots,y_N)$. Let $E(\vect{x})$ be the objective function, also referred to as {\em energy function}, defined by:
\begin{equation}\label{eq:obj_function}
\
E(\vect{x}) \defeq \alpha \sum_{n=1}^N x_ny_n + \sum_{m=1}^M \prod_{n\in H(m)} \!\!\! x_n, \,\ \forall \vect{x} \in \{-1, +1\}^N \!\!
\end{equation}
The first term of the objective function is the correlation between $\vect{x}$ and $\vect{y}$ (omitting the multiplicative coefficient $\alpha$), while the second term is the sum of the (bipolar) syndromes of $\vect{x}$. The multiplicative coefficient $\alpha > 0$ controls the contribution of the correlation term to the objective function. 
Clearly, the second term is maximized (equal to $M$) for any codeword $\vect{x}$. Hence, if the vector $\vect{x}$ maximizing $E(\vect{x})$ is a codeword, it is necessarily the ML decoding solution.  

The approach in \cite{wadayama2008gradient} is to maximize $E(\vect{x})$ (or minimize $-E(\vect{x})$) by using a variant of the gradient descent method, known as coordinate descent: an iterative algorithm that successively minimizes along coordinate directions ($x_n$). It turns out that the corresponding decoding algorithm is a variant of the BF decoding, referred to as Gradient Descent BF (GDBF). It consists of flipping the bits with lowest {\em local energy} values, where the local energy of a bit $x_n$, denoted by $E_n(\vect{x})$, is defined as:
\begin{equation}\label{eq:local_nrj}
E_n(\vect{x}) \defeq \alpha x_ny_n + \sum_{m\in H(n)} \prod_{n'\in H(m)} x_n'
\end{equation}
We shall simply denote $E_n \defeq E_n(\vect{x})$, when no confusion is possible. Thus, at each decoding iteration, the set of bit-flips is given by
\begin{equation}
{\cal F} \defeq \{ x_\nu \mid E_\nu \leq  E_{th} \},
\end{equation}
 where $E_{\text{th}}$ is a threshold value, referred to as {\em inversion threshold}. For the BSC model, it has been proposed in \cite{rasheed2014fault} to use $E_{\text{th}} = E_{\text{min}} \defeq \min_{n=1,\dots,N} E_n$, meaning that the set ${\cal F}$ contains the bits minimizing the local energy value (at each iteration). In this work, we use $E_{\text{th}} = E_{\text{min}} + \delta$, where $\delta$ is a predetermined value. We shall use $\delta=0$ for the BSC (or discrete-output channels), but $\delta > 0$ for the AWGN channel (or continuous-output channels).
%
According to the above description, GDBF is a {\em multi-bit flip} decoder, as multiple bits can be flipped at each iteration. 

 To escape from local minima, the PGDBF decoder \cite{rasheed2014fault} integrates a random perturbation of the bit-flip rule, consisting of flipping each bit $x_\nu \in {\cal F}$  with some probability $p < 1$ (the value of $p$ is determined empirically). The PGDBF decoder has been initially proposed and investigated for the BSC, but in this work we extend its use to both BSC and AWGN channels.

We note that for the AWGN channel, an alternative approach  is based on the NGDBF decoder \cite{sundararajan2014noisy},  which uses additive white Gaussian noise to perturb the local energy values (prior to determining the bit-flip set ${\cal F}$). 

\section{Randomized GDBF Decoders with Momentum}\label{sec:randomized_gdbf_with_momentum}
Gradient descent with momentum has first been introduced in \cite{nesterov27method}, as a method to accelerate the convergence rate of the gradient descent algorithm, and several extensions have been subsequently proposed in the literature, with different choices for the momentum parameter (see \cite{apidopoulos2018convergence} and references therein). Momentum is also used for stochastic gradient descent,  where the objective function is learned through samples in a training data set. 
It is one of the simplest extensions to gradient descent that has been successfully used for decades in the training of artificial neural networks \cite{zeiler2012adadelta}. 

Gradient descent with momentum remembers the update $\Delta x_n^{(\ell)} \defeq x_n^{(\ell)} -  x_n^{(\ell-1)}$ at each iteration $\ell$, and determines the next update as a linear combination of the gradient and the previous update.  This pushes the next update into the same direction as the previous update. 
Due to the discrete nature of our optimization problem (since $x_n^{(\ell)}\in\{\pm 1\}$), pushing in the same direction amounts to reducing the chances of a bit $x_n$ being flipped back again, after a bit-flip occurred. To do so, we add a {\em momentum term} to the local energy value of a bit $x_n$, as follows. 
\begin{equation}\label{eq:local_nrj_with_momentum}
E_n \defeq  \alpha x_ny_n + \sum_{m\in H(n)} \prod_{n'\in H(m)} x_n' + \rho(l_n),
\end{equation}
where $l_n\geq 1$ is equal to the number of iterations since the last flip of $x_n$ ($l_n=1$ if $x_n$ has been flipped at the previous iteration,  $l_n=2$ if $x_n$ has been flipped two iterations ago, and so on). We shall assume that the momentum {\em lasts for $L$ iterations}, meaning that $\rho(l_n) = 0$ if $l_n > L$. Hence, we define $\rho$ as a vector of $L$ non-increasing positive values:
\begin{equation}
\rho \ \defeq \ [\, \rho(1) \geq \rho(2) \geq \cdots \geq \rho(L) > 0 \,]
\end{equation}
and shall further set $\rho(L+1) \defeq 0$. Since there is no momentum when the decoding starts, $l_n$ values are initialized to $L+1$. The  $l_n$ value is incremented by $1$ at each iteration (without exceeding the maximum $L+1$ value), and set to $0$ each time the bit $x_n$ is flipped (such that it is incremented to $1$ the next iteration).

\begin{algorithm}[!b]
\caption{PGDBF decoding with momentum (w/M)}\label{alg:pgdbf_with_momentum}
\begin{algorithmic}[1]
{\small 

\smallskip
\Statex{\hspace*{-5mm}\parbox{10mm}{Input:}}  $\vect{y} = (y_1,\dots,y_N)\in{\mathbb R}^N$ \Comment{received word}
\Statex{\hspace*{-5mm}\parbox{10mm}{Output:}} $\vect{x} = (x_1,\dots,x_N)\in\{-1,+1\}^N$ \Comment{estimated codeword}  
\Statex{\hspace*{-5mm}\parbox{15mm}{Parameters:}} $\alpha>0$ (correlation coef.), $p\in\,]0,1]$ (bit-flip probability)
\Statex{\hspace*{-5mm}\parbox{15mm}{\ }} $\delta\geq 0$ (defines inversion threshold), $\rho$ (momentum)

\vspace*{-2mm}
\Statex \hspace*{-6mm}\rule{\linewidth+6mm}{0.5pt}

\ForAll{$n=1,\dots,N$} \Comment{\textbf{initialization}}
	\State $x_n = \text{sign}(y_n)$;
	\State $l_n = L+1$; 
\EndFor

\medskip
\ForAll{$\text{Iter} = 1,\dots, \text{Iter}_{\text{max}}$} \Comment{\textbf{iteration loop}}
	\MyFor{$m=1,\dots,M$} $c_m =  \prod_{n\in H(m)} x_n$; \Comment{syndrome}
	\EndMyFor
	
	\MyIf{$c_m = 1, \ \forall m=1,\dots,M$}  exit the iteration loop; 
	\EndMyIf
	
	\smallskip
	\ForAll{$n=1,\dots,N$} \Comment{local energy computation}
		\State $l_n = \min(l_n, L)+1$; 
   		\State $E_n = \alpha x_ny_n +  \sum_{m\in H(n)} c_m + \rho(l_n)$; 
   \EndFor
   
   \smallskip
   \State $E_{\text{th}} = \min_{n=1,\dots,N} E_n + \delta$; \Comment{inversion threshold}
   
   \smallskip
   \ForAll{$n=1,\dots,N$} \Comment{bit-flipping}
		\If{$E_n \leq E_{\text{th}}$ and $\text{rand}() < p$} 
			\State $x_n = -x_n$;
			\State $l_n = 0$;
		\EndIf 
   \EndFor
\EndFor
%
}
\end{algorithmic}
\end{algorithm}

The PGDBF decoding with momentum is described in Algorithm~\ref{alg:pgdbf_with_momentum} (according to our assumption, the BSC has input/output alphabet $\{-1,+1\}$). 
The parameters of the decoder are $\alpha > 0$ (the correlation coefficient),  $p\in\,]0,1]$ (the bit-flip probability), $\delta \geq 0$ (used to define the inversion threshold, which is set to $0$ for the BSC), and $\rho$ (the momentum). 
GDBF with momentum can be seen as a particular case, by taking the bit-flip probability parameter $p=1$. 

\begin{figure*}[!t]
\centering
\subfigure[$\ell=1$,  GDBF and GDBF-w/M]{\includegraphics[width=.22\linewidth]{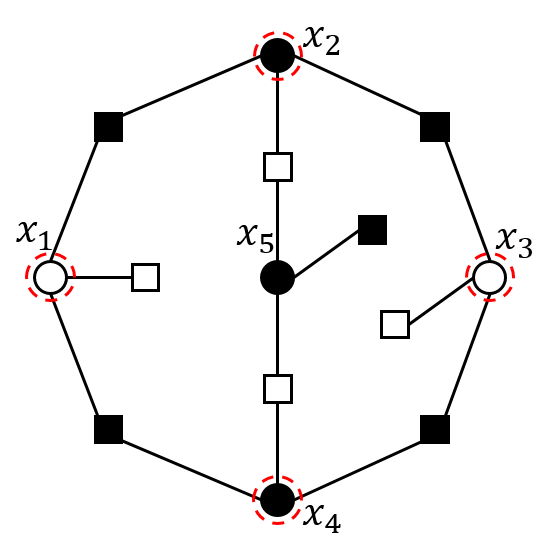}\label{fig:error_pattern_iter1}}\hfill%
\subfigure[$\ell=2$, GDBF]{\includegraphics[width=.22\linewidth]{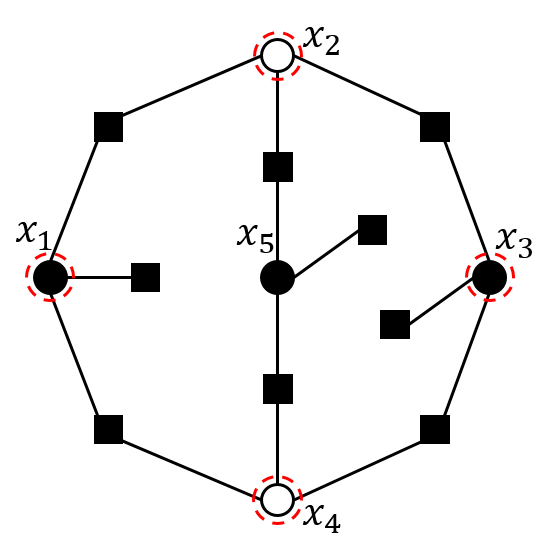}\label{fig:error_pattern_gdbf_iter2}}\hfill%
\subfigure[$\ell=2$, GDBF-w/M]{\includegraphics[width=.22\linewidth]{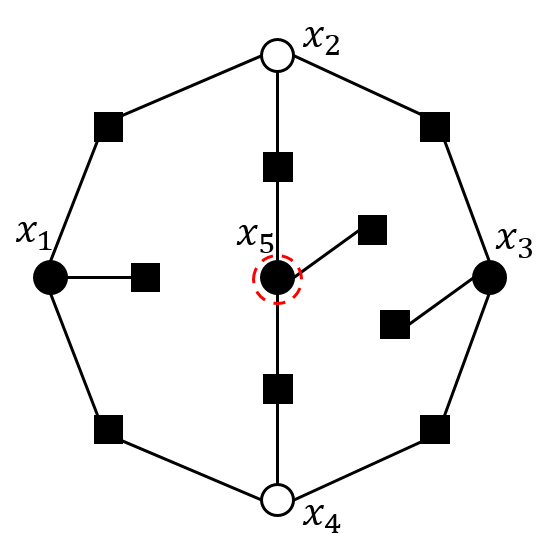}\label{fig:error_pattern_mgdbf_iter2}}\hfill%
\subfigure[$\ell=3$, GDBF-w/M]{\includegraphics[width=.22\linewidth]{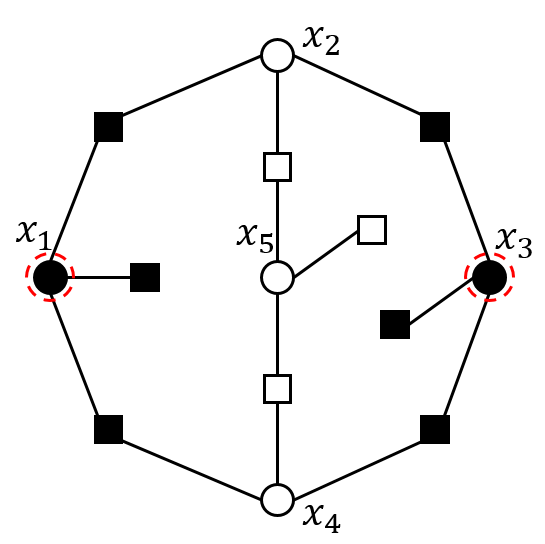}\label{fig:error_pattern_mgdbf_iter3}}
\caption{Trajectories of GDBF and GDBF-w/M decoders, for an error pattern corresponding to a (5,5) trapping set }
\label{fig:error_pattern}
\end{figure*}

\begin{figure*}
\centering
\includegraphics[width=.33\linewidth]{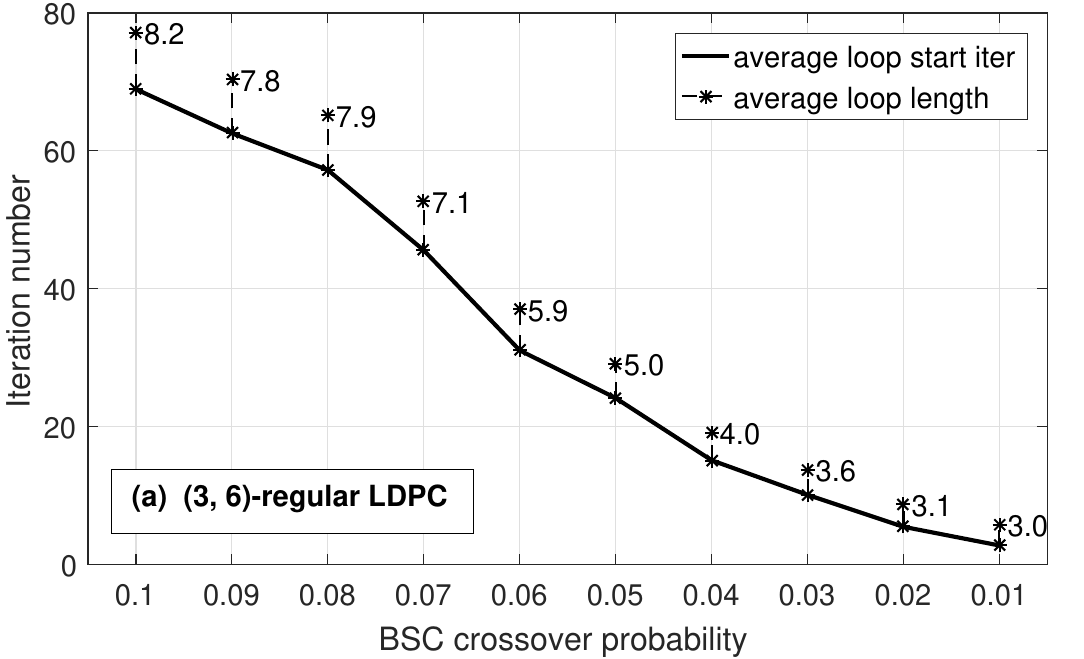}\hfill\includegraphics[width=.33\linewidth]{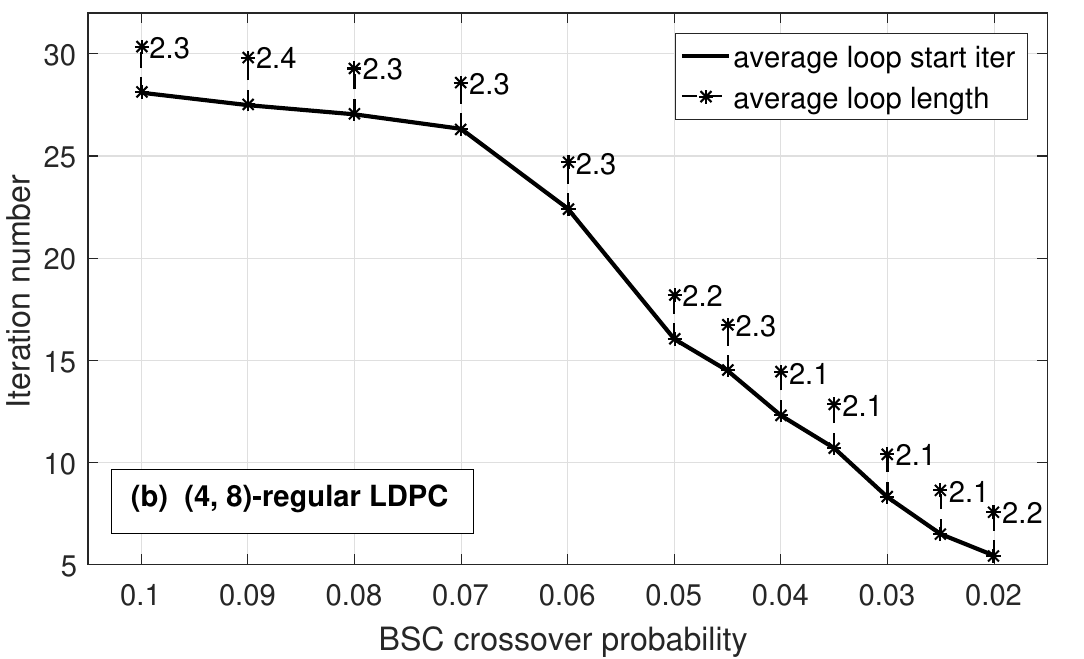}\hfill\includegraphics[width=.33\linewidth]{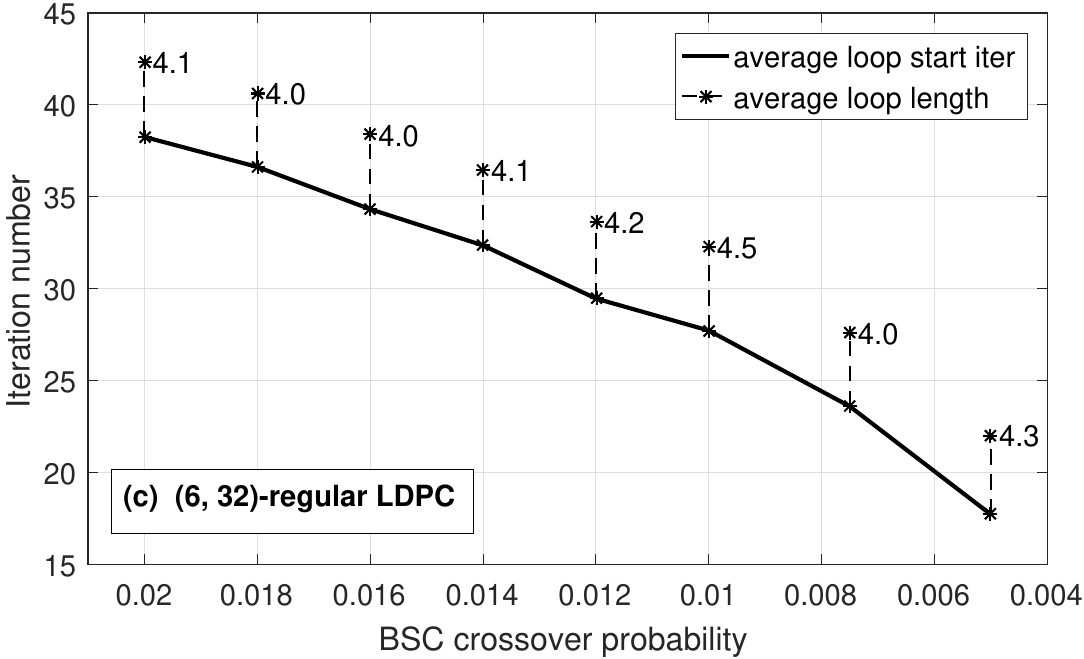} 
\caption{Average loop starting iteration and average loop length (regular LDPC codes, BSC)}
\label{fig:loop_detection}
\end{figure*}

\paragraph*{Escaping attractors} To illustrate the inertial effect of the GDBF-w/M decoder, we consider in Fig.~\ref{fig:error_pattern} the decoding of  an error pattern corresponding to a $(5,5)$ trapping set of an LDPC code with variable-nodes of degree $3$ (for a comparison with the PGDBF decoder, see  the similar analysis in \cite[Fig.~4]{rasheed2014fault}). Fig.~\ref{fig:error_pattern_iter1} illustrates the error pattern,
where variable-nodes are depicted as circles and check-nodes as squares, with full markers corresponding to a binary one (or bipolar $-1$, using our previous convention), and empty markers to binary zero (or bipolar $+1$). The iteration number is denoted as $\ell$. In the first iteration, variable-nodes $x_1,\cdots,x_4$ have minimum local energy  $E_n = 0$, $\forall n=1,\dots,4$. Therefore, these nodes are flipped in both GDBF and GDBF-w/M decoders, which is indicated in the figure by the surrounding  dashed red circles. In the second iteration, shown in Fig~\ref{fig:error_pattern_gdbf_iter2} for the GDBF decoder, the minimum local energy is obtained again for the same variable-nodes, since $E_n = -4$, $\forall n=1,\dots,4$. Hence, these nodes are flipped back, and the GDBF decoder cannot converge, as it will repeat again and again the same operations. Fig~\ref{fig:error_pattern_mgdbf_iter2} shows the second iteration for the GDBF-w/M decoder. In this case a momentum term is added to the local energy value of the bits that have been flipped in the first iteration. We will assume here that the momentum lasts only one iteration, and $\rho(1) = 3$. We get $E_n = -1$, $\forall n=1,\dots,4$, and the minimum local energy is obtained for $E_5 = -2$. Thus, the momentum term prevented the variable-nodes $x_1,\dots,x_4$ being flipped back again, and a new variable-node, namely $x_5$ is flipped in the second iteration. In the third iteration, shown in Fig~\ref{fig:error_pattern_mgdbf_iter3}, we get $E_1 = E_3 = -4$, $E_2 = E_4 = -2$, and $E_5 = 3$ (where we took  into account the momentum term added to variable-node $x_5$). Variable-nodes $x_1$ and $x_3$ are then flipped, leading to successful decoding.



\paragraph*{Momentum optimization}  
To determine the {\em length} ($L$) of the momentum, we start by investigating the behavior of the conventional GDBF decoder. Let us assume for the moment that the decoder is run for a possibly infinite number of iterations, and it only stops if  the syndrome check condition is satisfied, meaning that the estimated $\vect{x}$ is a codeword. We shall refer to the $\vect{x}$ vector as the {\em state} of the decoder. Clearly, the GDBF decoder is completely deterministic, and its state at some iteration only depends on its state at the previous iteration. Hence, assuming an infinite number of decoding iterations, the decoder will eventually reach either  a codeword state, or a state that has been visited before, in which case it gets caught in a loop, due to its deterministic behavior. If one keeps a history of the visited states, it is possible to detect when the decoder gets caught in a loop, {\em i.e.}, when $\vect{x}^{(\ell_1)} = \vect{x}^{(\ell_2)}$ for some iterations $\ell_1 < \ell_2$. We shall refer to $\ell_1$ as the loop starting iteration and to $\ell_2-\ell_1$ as the loop length. Fig.~\ref{fig:loop_detection} shows the average loop starting iteration and the average loop length, for three regular LDPC codes, over the BSC: 
\begin{itemize}
\item[(a)] $(3,6)$-regular LDPC code (rate = $0.5$), length $1296$ bits, 
\item[(b)] $(4,8)$-regular LDPC code (rate = $0.5$), length $1296$ bits, 
\item[(c)] $(6,32)$-regular LDPC code (rate = $0.84$), of length $2048$ bits, from the IEEE 802.3an standard. 
\end{itemize}
In all cases, the average loop starting iteration decreases with decreasing crossover probability of the BSC, meaning that the decoder gets caught in a loop earlier, as the channel gets better. The average loop length (indicated by the height of vertical bars) varies much less with the channel, especially for the higher degree codes: from  $2.3$ to $2.1$ for the $(4,8)$-regular code, and from  $4.08$ to $4.01$ for the $(6,32)$-regular code. 


Since the momentum is aimed at preventing the decoder getting caught in a loop, we use momentum of length $L$ equal (or close) to the average loop length of the GDBF decoder, as determined by simulation. 
%
Then, for a given moment length $L$, we search for momentum values $\rho_{max} \geq \rho(1) \geq \rho(2) \geq \cdots \geq \rho(L) > 0$, that yield the best decoding performance. To do this, we consider $\rho(\ell)$ in a discrete set of values, using some quantization step (to reduce the search space, we have used a quantization step to $0.5$). Although the above optimization process may be long and laborious, it is an offline optimization process, that needs to be performed only once, then the optimized momentum is integrated to the decoding algorithm. In order to avoid exhaustive search of all possible solutions, it might be advantageously combined with search heuristics based for instance on genetic algorithms.

%
%

\begin{table*}[!bp]
\centering
\caption{Parameters used by GDBF/PGDBF (w/M) decoders}
\label{tab:dec_parameters}
\begin{tabular}{|c|c|c|c|c|c|}
\hline
Channel & LDPC Code & correlation coef. $\alpha$ & inversion thresh. $\delta$ & bit-flip proba. $p$ & momentum $\rho$ \\
\hline
\hline
\multirow{3}{*}{BSC} & $(3,6)$-regular  & $0.5$ & $0$ & $0.9$ & $[2,2,2,1]$ \\ 
                     & $(4,8)$-regular  & $1.0$ & $0$ & $0.9$ & $[4,2,1]$ \\ 
                     & $(6,32)$-regular & $2.0$ & $0$ & $0.8$ & $[4,3,2,1]$\\ 
\hline
\hline
\multirow{2}{*}{AWGN} & $(4,8)$-regular  & $1.8$ & $1.1$ & $0.9$ & $[2, 2, 2, 2, 2, 1, 1]$ \\ 
                      & $(6,32)$-regular & $4.5$ & $1.2$ & $0.8$ & $[3, 3, 2, 1]$ \\ 
\hline
\hline
\multicolumn{6}{l}{Parameters used by each decoder: GDBF$(\alpha, \delta)$, PGDBF$(\alpha, \delta, p)$, GDBF-w/M$(\alpha, \delta, \rho)$, PGDBF-w/M$(\alpha, \delta, p, \rho)$}
\end{tabular}
\end{table*}

\begin{figure*}[!b]
\centering
\subfigure[(3,6)-regular LDPC]{\includegraphics[scale=.385]{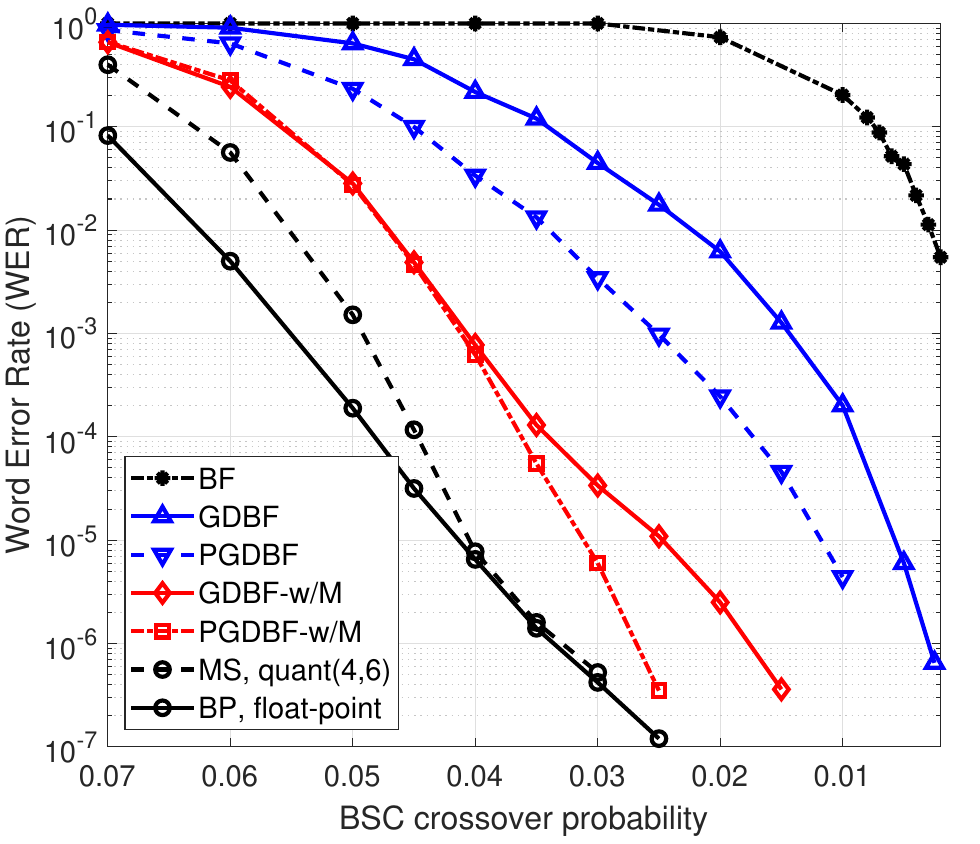}\label{fig:gdbf_with_momentum:iRISC_reg3x6}}\hfill%
\subfigure[(4,8)-regular LDPC]{\includegraphics[scale=.385]{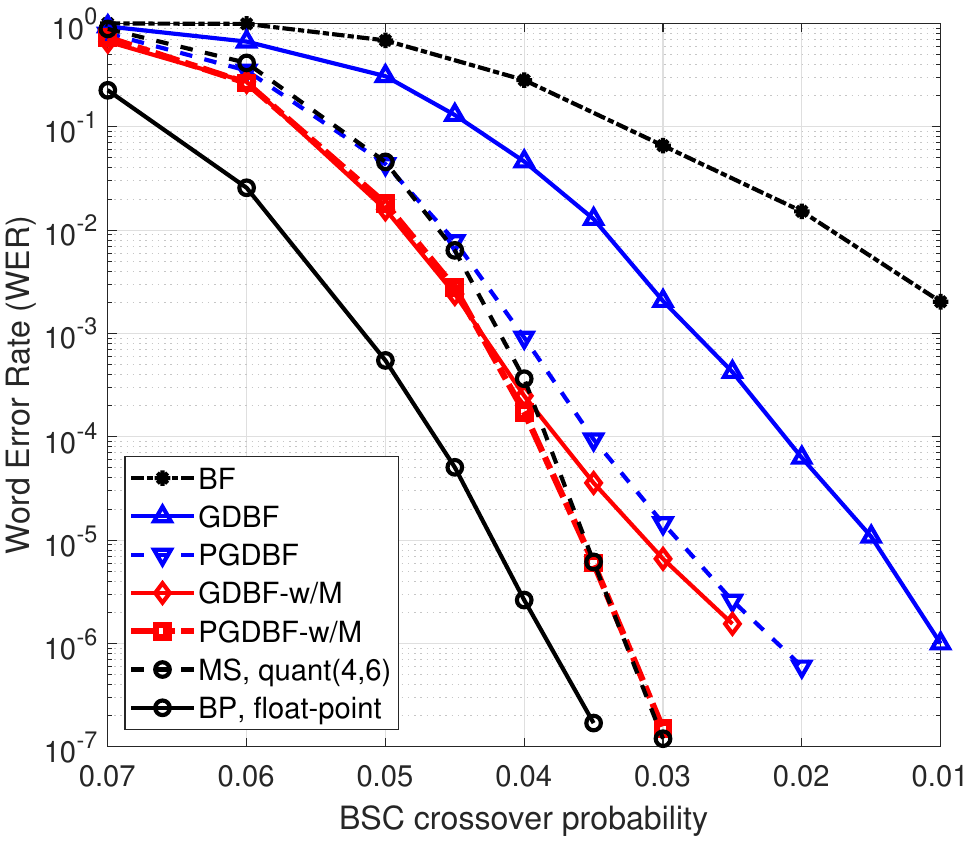}\label{fig:gdbf_with_momentum:iRISC_reg4x8}}\hfill%
\subfigure[(6,32)-regular LDPC]{\includegraphics[scale=.385]{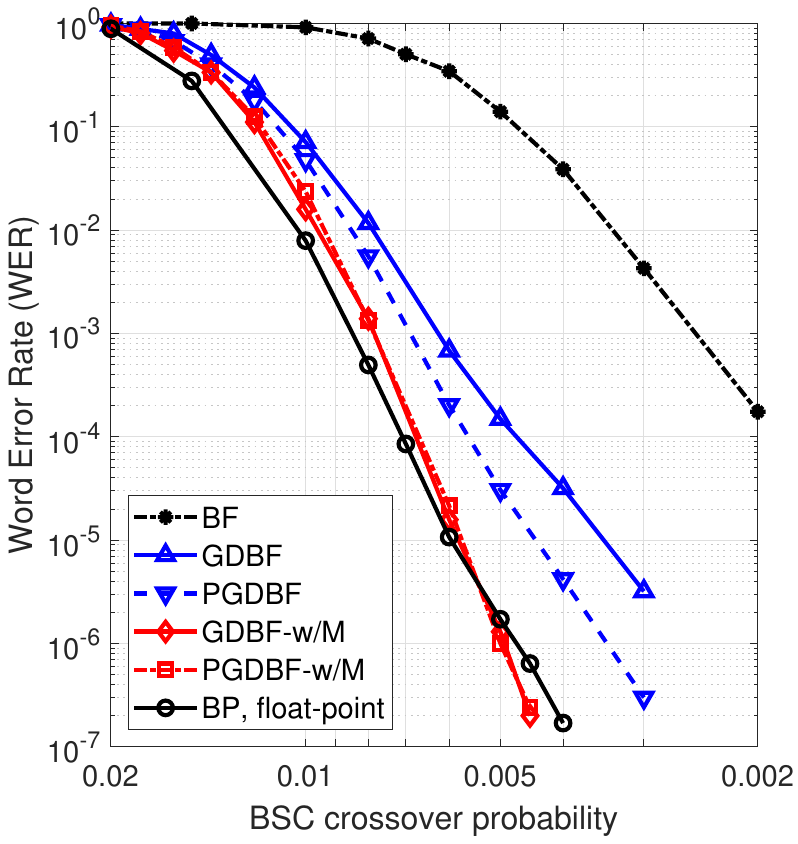}\label{fig:gdbf_with_momentum:ieee802_3_dv6}}  
\caption{Decoding performance for the BSC}
\label{fig:gdbf_with_momentum}
\end{figure*}

\section{Numerical Results}
In this section, the decoding performance of BF-based decoders (BF, GDBF/PGDBF, and GDBF/PGDBF with momentum (w/M)) is assessed against that of more powerful MP decoders (BP and MS). Table~\ref{tab:dec_parameters} summarizes the various parameters used by the GDBF/PGDBF (w/M) decoders. The list of parameters used by each decoder is also indicated in the table. In our simulations, MP decoders perform $50$ decoding iterations, with flooded scheduling, while BF-based decoders perform $300$ decoding iterations. 

\subsubsection*{BSC channel}
Simulation results for the BSC are shown in Fig.~\ref{fig:gdbf_with_momentum}. Regarding MP decoders, we consider the floating point BP decoder, and the finite-precision MS decoder, with 4-bit messages. 
%
%
For the {\em (3,6)-regular code}, momentum significantly improves the performance of both GDBF and PGDF decoders. It is worth noticing that GDBF-w/M decoder significantly outperforms PGDBF, although the former does not make use of randomness. The PGDBF-w/M decoder makes use of both randomness and momentum, closely approaching the performance of the floating-point BP and finite-precision MS decoders, especially in the error floor region. 
For the {\em (4,8)-regular code}, the GDBF-w/M decoder performs slightly better than PGDF, but they both exhibit an error floor at word error rate $\text{\sc wer} \approx 10^{-4}$.  The PGDBF-w/M decoder does not show any error floor down to $\text{\sc wer} = 10^{-7}$, and exhibits virtually the same decoding performance as the finite-precision MS decoder.
Finally, for the {\em (6,32)-regular code}, it can be seen that both GDBF-w/M and PGDBF-w/M decoders exhibit virtually the same decoding performance, closely approaching and even outperforming the floating-point BP decoding in the error floor region.

 \subsubsection*{AWGN channel} 
\begin{figure}[!t]
\centering
\subfigure[(4,8)-regular LDPC]{\includegraphics[width=.9\linewidth]{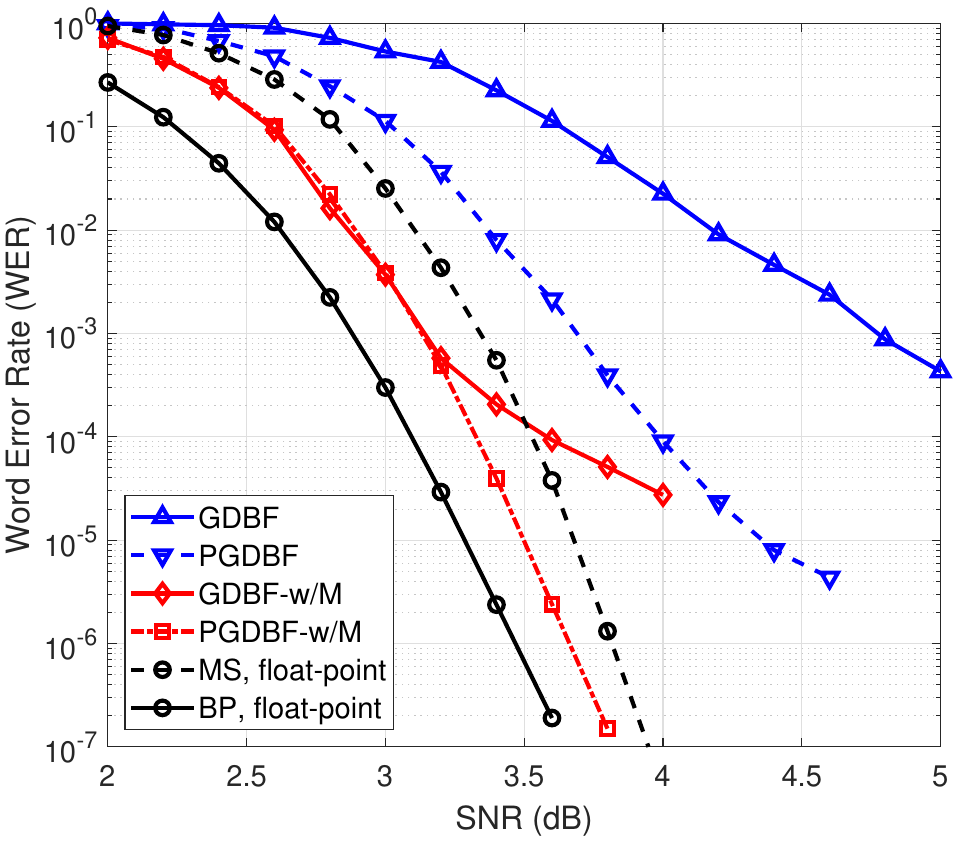}\label{fig:gdbf_with_momentum_awgn:iRISC_reg4x8}}
\subfigure[(6,32)-regular LDPC]{\includegraphics[width=.9\linewidth]{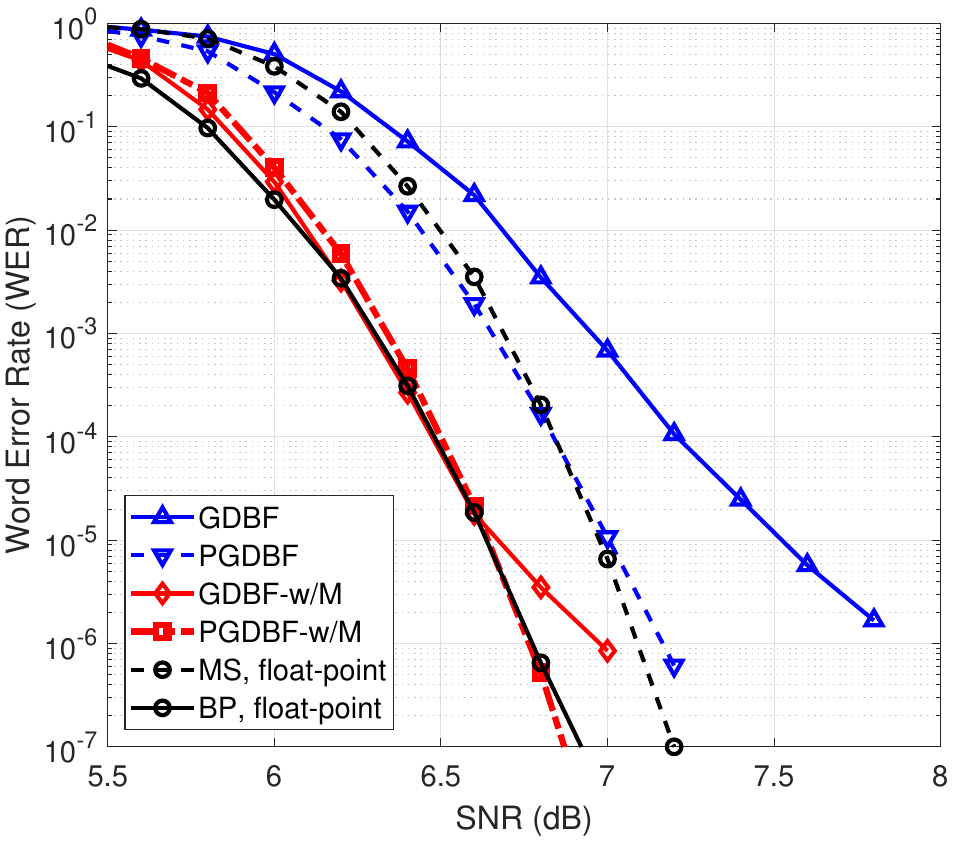}\label{fig:gdbf_with_momentum_awgn:ieee802_3_dv6}}  
\caption{Decoding performance for the AWGN channel}
\label{fig:gdbf_with_momentum_awgn}
\end{figure}

Simulation results for the AWGN channel (with bipolar $\pm 1$ inputs) are shown in Fig.~\ref{fig:gdbf_with_momentum_awgn}. Both BP and MS decoders are implemented in floating-point precision. GDBF-w/M and PGDBF-w/M decoders show virtually the same decoding performance in the waterfall region, outperforming the floating-point MS decoder, but the GDBF-w/M decoder exhibits a higher error floor. The PGDBF-w/M decoder closely approaches (within less than $0.2$\,dB for the (4,8)-regular code) or achieves virtually the same performance (for the (6,32)-regular code) as the floating-point BP decoder. This is all the more remarkable for soft-output channels, and demonstrates the effectiveness of the momentum technique for avoiding local minima in gradient descent based decoding.

\section{Conclusion and Perspectives}

This paper reported on a new approach aimed at improving the decoding performance of randomized GDBF decoders, inspired by the momentum technique used in gradient descent optimization. We showed that GDBF or randomized GDBF decoders with momentum may closely approach the floating-point BP decoding performance, and may even outperform it in the error-floor region, especially for graphs with high connectivity degree. This makes the proposed technique particularly relevant to low error-floor applications, since LDPC codes designed for such applications are usually defined by bipartite graphs with higher degree of connectivity.

The results presented in this paper open a number of perspectives, 
regarding the optimization of the momentum parameters, for a given bipartite graph (or the optimization of both randomization and momentum parameters, in case of randomized GDBF decoders). To optimize the decoding performance in the error floor region, an analytical model based on absorbing Markov
chains could be used to quantify the contribution of dominant trapping/absorbing 
sets to the word error rate~\cite{ivanivs2016error}. 
%
 A different approach is to learn optimal decoding parameters, by using deep reinforcement learning techniques.
BF decoding {\em de facto} behaves as a deep neural network (DNN), with input/output layers corresponding to the input/output  of the decoder, and hidden layers corresponding to the decoding iterations. We believe that DNN models for GDBF-based decoders may be used to learn decoding parameters ({\em e.g.}, randomization and momentum parameters), but also to learn weights and bias to improve decoding performance, while taking into account the specific code structure (short cycles, trapping/absorbing sets, etc.). 

\bibliographystyle{IEEEtran}
\bibliography{./biblio/biblio_database}

%
%
%
%




\pagebreak

\onecolumn
\appendices

\section{Additional Data}

The $(3,6)$-regular and $(4,8)$-regular LDPC codes used for the simulations in this paper are quasi-cyclic LDPC codes, with base matrix of size $12 \times 24$ and expansion factor $z = 54$. Hence, both codes have length $24\times 54 = 1296$ bits. The base matrices of the two codes are given below. The parity check matrix of a code is obtained by replacing each non-zero entry $b\geq 0$ in the base matrix by a square matrix of size $z\times z$, defined as the circular {\em right-shift} of the identity matrix by $b$ positions. Entries $b=-1$ in the base matrix are replaced by all-zero square matrices of size $z\times z$.

\medskip
\noindent $(a)$ $(3,6)$-regular LDPC code (rate $=0.5$), length $1296$ bits

\smallskip\noindent
{\footnotesize $\left[\begin{array}{*{24}{c}}
 49 & \m & \m & \m & \m & 43 & \m & \m & \m & \m & 50 & \m & \m & \m & \m &  2 & \m & 27 & \m & \m & \m & \m & \m & 49 \\
 \m & \m & \m & 10 & 41 & \m & \m & \m & \m & 52 & \m & \m & 32 & \m & \m & \m & \m & \m & 50 & \m & 50 & \m & \m & \m \\
 \m & \m & 20 & \m & \m & \m & \m & 20 & \m & \m & \m & 51 & \m & 10 & \m & \m & 47 & \m & \m & \m & \m & \m & 33 & \m \\
 \m & 24 & \m & \m & \m & \m & 22 & \m & 53 & \m & \m & \m & \m & \m & 31 & \m & \m & \m & \m & 18 & \m & 47 & \m & \m \\
 10 & \m & \m & \m & 15 & \m & \m & \m & \m & \m &  2 & \m & \m & \m & \m & 50 & \m & 13 & \m & \m & \m & \m & \m & 53 \\
 \m & \m & 44 & \m & \m &  6 & \m & \m & \m & \m & \m & 29 & \m & 40 & \m & \m & 16 & \m & \m & \m & 13 & \m & \m & \m \\
 \m &  2 & \m & \m & \m & \m & \m & 13 & 41 & \m & \m & \m & \m & \m & 42 & \m & \m & \m & \m & 48 & \m & 49 & \m & \m \\
 \m & \m & \m & 36 & \m & \m & 24 & \m & \m & 50 & \m & \m & 12 & \m & \m & \m & \m & \m & 10 & \m & \m & \m & 48 & \m \\
 \m & \m & 47 & \m & 50 & \m & \m & \m & \m & \m &  0 & \m & \m & \m & \m &  9 & \m &  7 & \m & \m & \m & \m & \m & 28 \\
 \m & 24 & \m & \m & \m & \m & \m & 51 & \m & 38 & \m & \m & \m & \m &  6 & \m & \m & \m & \m & 23 & \m & 16 & \m & \m \\
  6 & \m & \m & \m & \m & \m &  5 & \m & \m & \m & \m & 13 & \m &  3 & \m & \m & 29 & \m & \m & \m & 16 & \m & \m & \m \\
 \m & \m & \m & 35 & \m & 16 & \m & \m & 37 & \m & \m & \m &  4 & \m & \m & \m & \m & \m & 24 & \m & \m & \m & 29 & \m
\end{array}\right]$}

\bigskip 
\noindent $(b)$ $(4,8)$-regular LDPC code (rate $=0.5$), length $1296$ bits

\smallskip\noindent
{\footnotesize $\left[\begin{array}{*{24}{c}}
 11 & \m & \m & \m & 27 & \m & \m & \m & 33 & 16 & \m & \m & \m & 44 & \m & \m & 44 & \m &  8 & \m & \m & \m & \m &  0 \\
 \m & 25 & \m & \m & \m & 31 & 29 & \m & \m & \m & 29 & \m & \m & \m & 36 & \m & \m & 34 & \m & 15 & \m & \m & 17 & \m \\
 \m & \m & 44 &  4 & \m & \m & \m & 11 & \m & \m & \m &  2 & 50 & \m & \m & 52 & \m & \m & \m & \m & 30 & 33 & \m & \m \\
 27 & \m & \m & \m & 34 & \m & 20 & \m & \m & 20 & \m & \m & \m & 13 & \m & \m & 27 & \m &  4 & \m & \m & \m & \m & 27 \\
 \m & 42 & \m & 22 & \m & \m & \m & 11 & \m & \m & \m & 44 & \m & \m &  4 & 14 & \m & \m & \m & \m & 45 & 17 & \m & \m \\
 \m & \m & 24 & \m & \m & 10 & \m & \m & 10 & \m & 18 & \m &  2 & \m & \m & \m & \m & 19 & \m & 38 & \m & \m & 31 & \m \\
 \m & \m & 40 & \m & \m & 35 & \m & \m & 31 & 19 & \m & \m &  3 & \m & \m & 42 & \m & \m & \m & 42 & \m & \m & 39 & \m \\
 \m & 29 & \m &  0 & \m & \m & \m & 29 & \m & \m &  5 & \m & \m & \m & 47 & \m & \m & 28 & \m & \m & 28 & 41 & \m & \m \\
  9 & \m & \m & \m &  7 & \m & 20 & \m & \m & \m & \m &  1 & \m & 19 & \m & \m &  5 & \m & 25 & \m & \m & \m & \m & 41 \\
 \m & \m & 53 & \m & \m &  3 & \m & \m & 26 & \m &  3 & \m & \m & \m & 30 & \m & \m &  5 & \m & 35 & \m & \m & 44 & \m \\
 \m &  4 & \m & \m &  4 & \m & \m &  5 & \m & \m & \m & 13 & 42 & \m & \m & 50 & \m & \m & \m & \m & 36 & 38 & \m & \m \\
 39 & \m & \m & 17 & \m & \m & 36 & \m & \m & 34 & \m & \m & \m & 46 & \m & \m & 12 & \m &  8 & \m & \m & \m & \m & 15
\end{array}\right]$} 

\end{document}